\documentclass[12pt]{article}
\usepackage{wrapfig,lipsum,booktabs}
\usepackage{multicol}
\usepackage{times}
\usepackage{geometry}
\usepackage[utf8]{inputenc}
\usepackage{enumitem,amssymb}
\usepackage[justify]{ragged2e}
\newlist{thematic}{itemize}{8}
\setlist[thematic]{label=$\square$}
\usepackage{pifont}
\usepackage{graphicx}
 \usepackage[numbers]{natbib}
 \usepackage{wasysym}

 \usepackage{color}
\usepackage[usenames,dvipsnames,svgnames,table]{xcolor}
\usepackage[colorlinks=true,
            linkcolor=blue,
            urlcolor=blue,
            citecolor=blue]{hyperref}

\begin{document}
\setlength{\textfloatsep}{8pt plus 1.0pt minus 2.0pt}

\raggedright
\huge
Astro2020 Science White Paper \linebreak

Brown Dwarfs and Directly Imaged Exoplanets in Young Associations\linebreak
\normalsize

\noindent \textbf{Thematic Areas:} \hspace*{55pt} $\XBox$ Planetary Systems \hspace*{14pt} $\XBox$ Star and Planet Formation \hspace*{20pt}\linebreak
$\Square$ Formation and Evolution of Compact Objects \hspace*{31pt} $\Square$ Cosmology and Fundamental Physics \linebreak
  $\XBox$  Stars and Stellar Evolution \hspace*{1pt} $\Square$ Resolved Stellar Populations and their Environments \hspace*{40pt} \linebreak
  $\Square$    Galaxy Evolution   \hspace*{45pt} $\Square$             Multi-Messenger Astronomy and Astrophysics \hspace*{65pt} \linebreak
  
Name:	Jacqueline K. Faherty
 \linebreak						
Institution:  American Museum of Natural History
 \linebreak
Email: jfaherty@amnh.org
 \linebreak
Phone:  +1 (212) 496-3527
 \linebreak
 
\textbf{Co-authors:}  \\
Katelyn Allers ~(Bucknell University)\\
Daniella Bardalez Gagliuffi (American Museum of Natural History)\\
Adam J. Burgasser (University of California San Diego)\\
Jonathan Gagn\'e (Universit\'e de Montr\'eal) \\
John Gizis (University of Delaware)\\
J. Davy Kirkpatrick (IPAC)\\
Adric Riedel (STSCI)\\
Adam Schneider (Arizona State University)\\
Johanna Vos (American Museum of Natural History)\\

\vskip 0.5in

\begin{justify}
\textbf{Abstract:}
In order to understand the atmospheres as well as the formation mechanism of giant planets formed outside our solar system, the next decade will require an investment in studies of isolated young brown dwarfs. In this white paper we summarize the opportunity for discovery space in the coming decade of isolated brown dwarfs with planetary masses in young stellar associations within 150 pc. We suggest that next generation telescopes and beyond need to invest in characterizing young brown dwarfs in order to fully understand the atmospheres of sibling directly imaged exoplanets as well as the tail end of the star formation process. 
\end{justify} 
\pagebreak
\begin{wraptable}{!r}{8.0cm}
\footnotesize
\caption{\bf Young Groups Found within 150pc of the Sun defined by BANYAN $\Sigma$\label{tab:younggroups}}
\begin{tabular}{cccc}
Banyan Name &  Full Name & Age (Myr)  \\
\hline \\
ABDMG	& AB Doradus	& 149$^{+51}_{-19}$\\
\hline
\hline
BPMG	& $\beta$ Pictoris	& 24$\pm$3\\
\hline
\hline
CAR	& Carina	&	 45$^{+11}_{-7}$\\
\hline
\hline
CARN	& Carina-Near		&$\sim$200 \\
\hline
\hline 
CBER	&Coma Berenices 	&562$^{+98}_{-84}$\\
\hline
\hline 
COL	&	Columba	&	42$^{+6}_{-4}$	\\
\hline
\hline 
EPSC	&	$\epsilon$ Chamaeleontis	&3.7$^{+4.6}_{-1.4}$\\
\hline
\hline 
ETAC	&$\eta$ Chamaeleontis	&11$\pm$3\\
\hline
\hline 
HYA	& the Hyades cluster &750$\pm$100	\\
\hline
\hline 
IC2602	&IC 2602 &46$^{+6}_{-5}$	\\
\hline
\hline 
LCC	&	Lower Centaurus Crux		&15$\pm$3\\
\hline
\hline 
OCT	&	Octans 	&35$\pm$5	\\
\hline
\hline 
PL8	&	Platais 8 	&$\sim$60\\
\hline
\hline 
PLE	&	the Pleiades cluster	&112$\pm$5\\
\hline
\hline 
THA	&	Tucana-Horologium 		&45$\pm$4\\
\hline
\hline 
THOR	&	32 Orionis 	&22$^{+4}_{-3}$\\
\hline
\hline 
TWA	&	TW Hydra 	&10$\pm$3\\
\hline
\hline 
UCL	&Upper Centaurus Lupus&16$\pm$2	\\
\hline
\hline 
UMA	&	Ursa Major cluster &414$\pm$23	\\
\hline
\hline 
USCO	&	Upper Scorpius	&10$\pm$3\\
\hline
\hline 
XFOR	&	$\chi$ Fornax	&$\sim$500\\
\end{tabular}
\end{wraptable} 

\begin{justify}
\begin{figure}[h]
 \begin{center}
 \includegraphics[scale=.35]{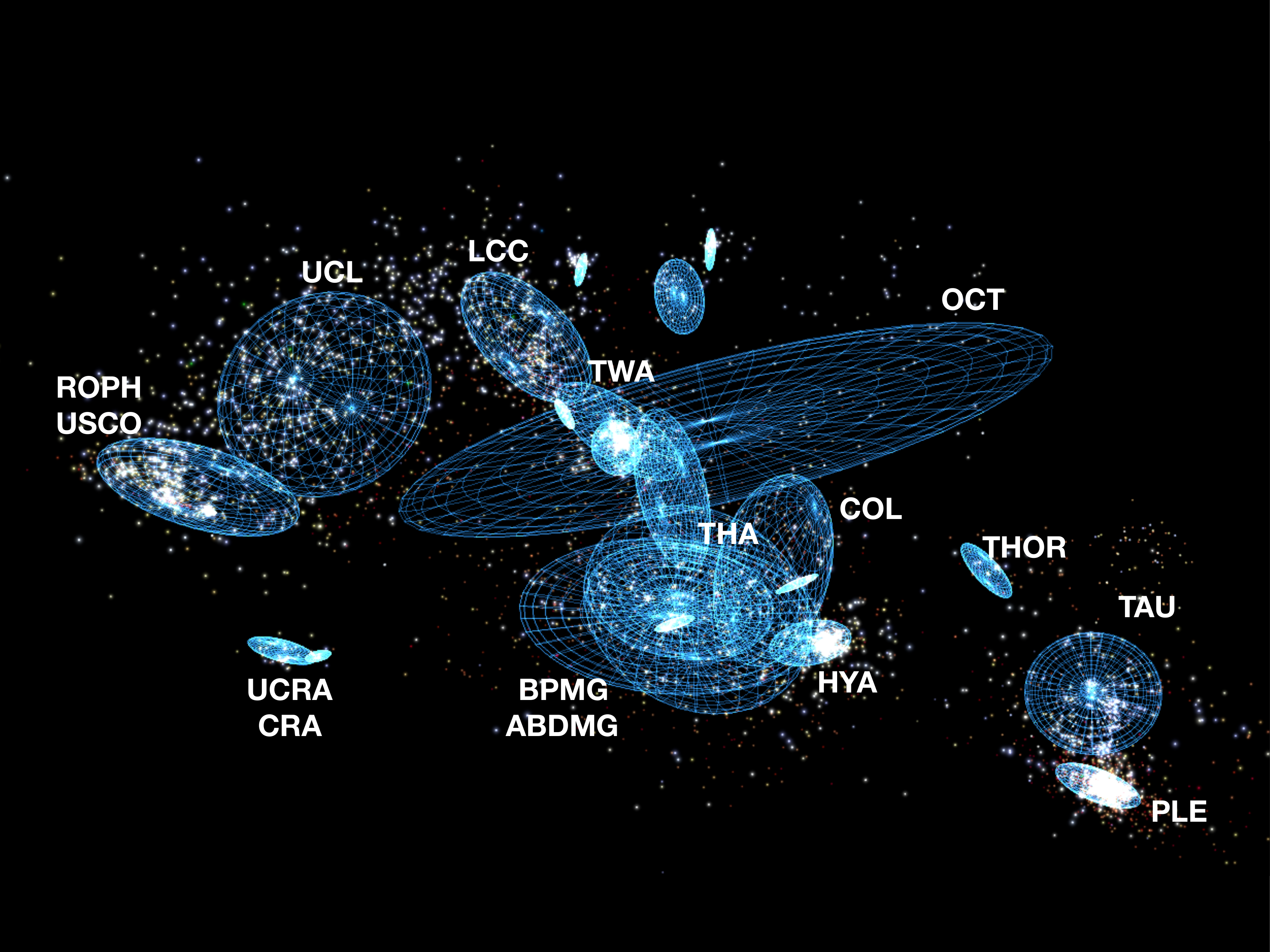}
\end{center}
\caption{\footnotesize An image of the 150 pc area around the Sun highlighting the locations of star forming regions and nearby moving groups.  Table~\ref{tab:younggroups} describes each association.}
 \label{fig:groups}
\end{figure}

\begin{center}
\vskip -0.2in
{\bf 1. Overview on the Importance of Young Associations} 
\end{center}
\vskip -0.1in
Young stellar associations are a goldmine for studies of the tail end of the mass function. Within 150 pc of the Sun there are more than 20 known associations of young stars known to be co-moving, of a similar age, and densely populated in space (see Figure ~\ref{fig:groups} and Table ~\ref{tab:younggroups} for a summary as well as \citep{Kastner19a}).  In-depth studies of the closest associations to the Sun have revealed that they harbor large numbers of low–mass stars, brown dwarfs, and even objects whose mass falls below the deuterium burning boundary (so called free-floating planetary–mass objects; e.g., \citep{Liu13}, \citep{Gagne18}, \citep{Faherty16}). Moving groups contain the closest young stars to the Sun, therefore they are also the targeting ground of direct imaging campaigns for warm exoplanets. Associations such as Tucana-Horologium, TW Hydra, $\beta$ Pictoris and the AB Doradus moving group contain isolated objects that range in mass from a few solar masses down to a few Jupiter masses as well as stars with planetary–mass companions (\citep{Macintosh15}, \citep{Lagrange10}, \citep{Artigau15}, \citep{Schneider16}, \citep{Kellogg16}, \citep{Naud14}). Observations of these associations enable investigations of the mass function, kinematics, and spatial distribution across the full range of objects generated through star formation processes in different isolated groups at young ($\sim$1–2 Myr), medium ($\sim$30–50 Myr), and older ($\sim$100–700 Myr) ages.
\medskip\noindent


In the coming decade, ESA's Gaia mission \citep{GaiaDR2} will completely revolutionize our understanding of these nearby coherent groups. The vast number of stars within 150 pc with well measured kinematics will lead to greatly expanded membership lists of known associations.  The microarcsecond precision of Gaia parallaxes and proper motions will allow us to more finely differentiate between groups with overlapping spatial positions and velocity dispersions and will likely lead to the discovery of entirely new and previously overlooked associations. Indeed, an examination of the Tycho Gaia Astrometric Survey (TGAS) as well as Gaia DR2 has already led to the refinement of known groups, expanded membership lists and the establishment of new associations (e.g. \citep{Gagne18b}, \citep{Faherty18}, \citep{Gagne18c}, \citep{Oh17}, \citep{Luhman18}, \citep{Tang19}). However, despite Gaia's ability to provide precise astrometry for young brown dwarfs in these associations, it cannot reach the expected bottom of the mass function at even the closest group at the youngest age. 
\medskip\noindent

\begin{figure}[!h]
 \begin{center}
 \includegraphics[scale=0.35]{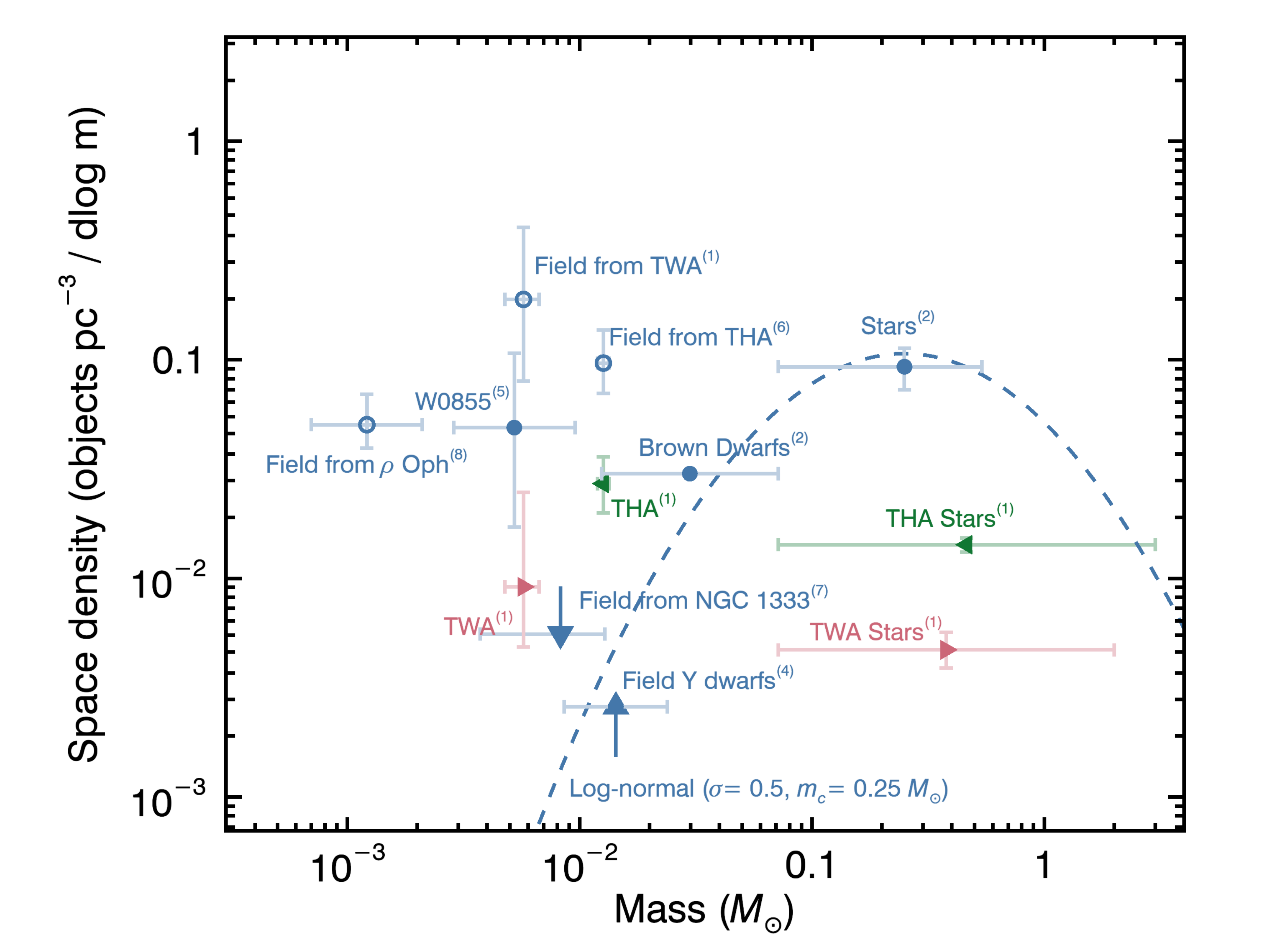}
\end{center}
\vskip -0.2in
\caption{\footnotesize A plot of space density estimates from various surveys compiled by \citep{Gagne17b}. The dashed line shows a fiducial log-normal IMF. This figure demonstrates how the current planetary-mass space density estimates of Tucana-Horologium and TW Hydra are higher than the predictions from a typical log-normal IMF anchored on the space density of main-sequence stars in the field, even though the stellar densities of both associations are much sparser than those of field stars. As powerfully demonstrated in Figure~\ref{fig:combo}, a CatWISE and SPHEREx search will fill in the basement of the mass function for many more groups.
}
 \label{fig:spacedensity}
\end{figure}

\medskip\noindent

\begin{center}
\vskip -0.3in
{\bf 2. The Initial Mass Function Into the Planetary Mass Regime.} 
\end{center}
Initial mass function studies of nearby star-forming regions have begun probing the basement of the mass function (see Figure~\ref{fig:spacedensity}). While Gaia has provided (and will continue to do so in the coming decade) some higher-mass brown dwarf candidates, one must look in the infrared in order to fill in the lowest temperature young objects near the Sun (see also for example white papers from \citep{Leggett19a}, \citep{Kirkpatrick19a}).  The Wide-field Infrared Survey Explorer mission (WISE; \citep{Wright10}), the Two Micron All Sky Survey (2MASS; \citep{Skrutskie06}), and Pan-Starrs (\citep{Magnier13}) have all played a significant role in young brown dwarf studies. WISE in particular has identified candidates in star-forming groups such as Chamaeleon I, Taurus, and Perseus (\citep{Esplin17b}; \citep{Esplin17}). While these very young associations ($\sim$1–3 Myr) and their yield of free floating planet candidates are exciting, the groups suffer from (at times extreme) reddening and are at larger distances making their lowest mass members difficult to study in detail.  
\medskip\noindent
\end{justify}

Nearby ($<$ 100 pc) and slightly older ($\sim$10-50 Myr) young associations are not reddened and provide the opportunity for an easier characterization, but identifying the members of these nearer associations is more challenging because they are spread over large fractions of the sky therefore they require observationally expensive parallaxes and radial velocities to confirm membership. A dozen planetary-mass objects (model-dependent masses; see \citep{Burgasser19a} and \citep{BardalezGagliuffi19a} white papers for alternate mass determination methods) in the range 6–13 M$_{Jup}$ have recently been detected, making it possible to obtain spatial densities down to the lowest-masses of star formation in nearby young moving groups (see Figure ~\ref{fig:spacedensity}). These first results suggest an over-density of planetary-mass objects compared to the predictions of a fiducial log-normal initial mass function, but they are still based on small-number statistics and are inconsistent with some open cluster studies (\citep{Scholz12}; \citep{Muzic17}).  As stated below, the coming decade will see a push into discoveries in large numbers of the lowest possible masses formed through the star formation process but these new objects will require next generation instruments for membership confirmation and necessary characterization studies. 
\medskip\noindent

\begin{justify}
\begin{figure}[!h]
 \begin{center}
 \includegraphics[scale=.45]{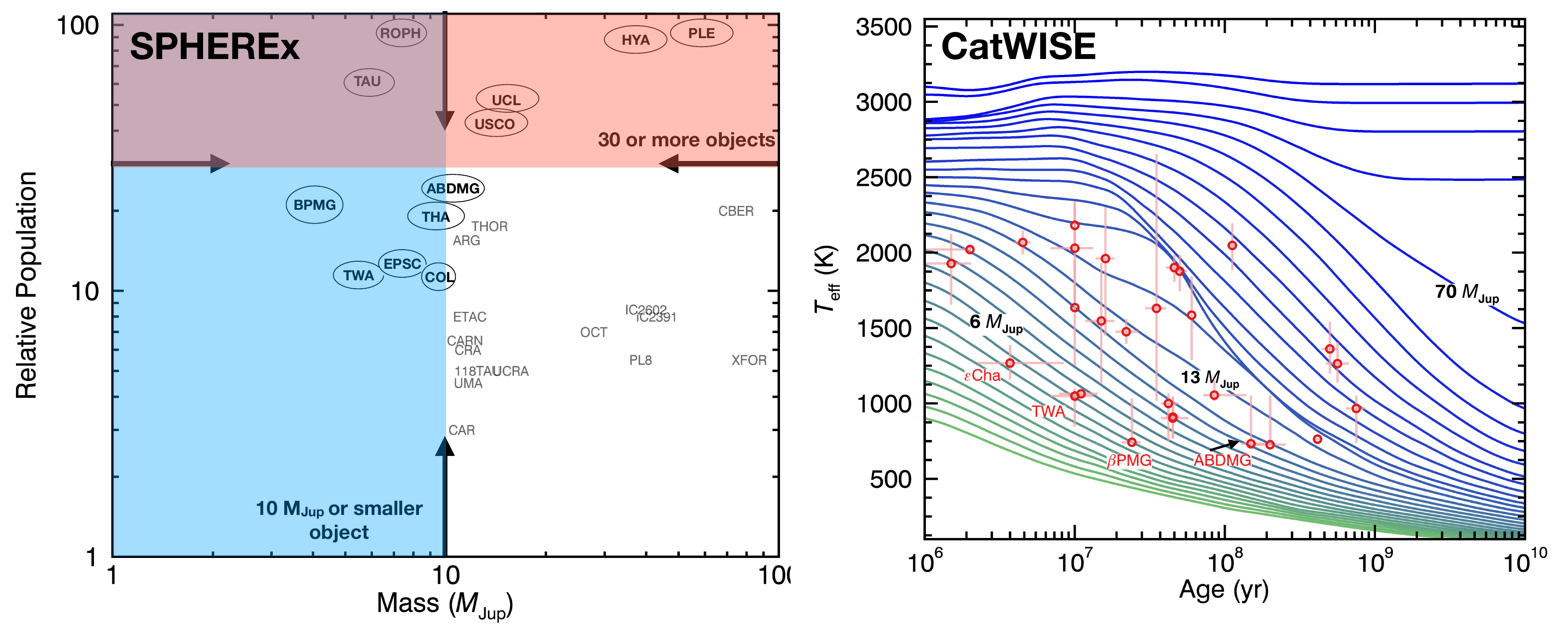}
\end{center}
\vskip -0.1in
\caption{\footnotesize Left panel:  Using the limiting magnitude of SPHEREx with the distances to each of the young associations within 150 pc of the Sun (see Table ~\ref{tab:younggroups} we estimate how many low mass objects SPHEREx will observe.  Right panel:  Using the limiting magnitude and proper motion uncertainty of CatWISE with the known kinematics of young associations we estimate the lowest mass object detectable in each group.}
 \label{fig:combo}
\end{figure}

\begin{center}
\vskip -0.3in
{\bf 3. Discoveries of Young Isolated Planetary Mass Objects In the Next Decade} 
\end{center}
The majority of young brown dwarf discoveries to date have been made using a combination of 2MASS, WISE, and/or Pan-STARRs data (e.g. \citep{Best17}, \citep{Aller16}, \citep{Schneider17}, \citep{Gagne15}, \citep{Faherty13}).  However the limiting magnitudes and proper motions by the individual catalogs or the combination of catalogs meant that we have been restricted in discovery space for the lowest mass isolated objects.  At present, there remains a large amount of kinematic discovery space for WISE, in particular with new all-sky epochs emerging from the ongoing NEOWISE campaign, which provides a 5-year baseline for higher precision and lower value proper motions (\citep{Mainzer14}). The coming decade will see astronomers taking advantage of that unexplored data.  One catalog that will soon come on line is an all-sky kinematic catalog named CatWISE (Peter Eisenhardt NASA ADAP) with moderate precision proper motions for millions of faint (W2 $<$ 15.8 magnitude) mid-infrared sources.  Using CatWISE as an example of the discoveries in waiting, we have investigated the potential of WISE data for uncovering new low temperature members of known moving groups and star forming regions. The right hand panel of Figure ~\ref{fig:combo} shows the evolutionary models from \citep{Saumon08} overplotted with the lowest-mass objects that can be detected by a catalog like CatWISE in the 21 closest young associations to the Sun (listed in Table ~\ref{tab:younggroups}). In groups such as AB Doradus, $\beta$ pictoris, $\eta$ Cha and TW Hya, we have the capability of discovering isolated objects with masses down to 6 M$_{Jup}$. 
\medskip\noindent

In addition to new WISE discoveries that we expect in the coming decade, there is the potential of discovery and characterization by NASA's recently selected mission called the Spectro-Photometer for the History of the Universe, Epoch of Re-ionization and Ices Explorer (SPHEREx) mission slated to launch in 2023.  SPHEREx is an all-sky mission that will take low resolution spectra in the near to mid infrared of millions of sources (within a 6'' pixel radius) and will be very impactful on young brown dwarf studies.  The left hand panel in Figure ~\ref{fig:combo} uses the stated limiting magnitudes that SPHEREx will be operating with along with the distance boundaries for the young associations listed in Table ~\ref{tab:younggroups} to demonstrate what we expect the yield of low resolution, near- to mid-infrared spectra will be in each group.  As shown, a combination of the CatWISE catalog to kinematically identify sources and SPHEREx to obtain follow-up spectra will lead to an unprecedented library of very low mass sources.  Several groups such as $\beta$ Pictoris, TW Hydra, and Taurus will be examined for an abundance of objects $<$ 10 M$_{Jup}$. 
\medskip\noindent

While this is extremely exciting to envision, one striking problem remains: in order to confirm membership in a young association we require high precision infrared parallaxes and radial velocities.  While ground-based parallax programs have been tremendously effective at surveying brown dwarfs (e.g. \citep{Dupuy12}, \citep{Liu16}, \citep{Faherty12}, \citep{Vrba04}, \citep{Tinney14}), the infrared magnitudes of isolated planetary-mass objects at the distances of young associations requires a new generation of instruments.  As discussed in the white paper by \citep{Kirkpatrick19a}, the \textit{James Webb Space Telescope (JWST)} is ill suited for these astrometric studies because of high overheads in slewing between targets.  WFIRST, with its 18-array infrared focal plane might be a suitable astrometric imager and would be greatly enhanced if the simple K-band filter was installed (see for example the white paper by \citep{Stauffer19a}).
\medskip\noindent

Moreover, the low resolution spectra of SPHEREx will help fill in the bolometric luminosities of new discoveries but will not allow us to probe atmosphere, metallicity, or gravity parameters in detail.  For a more detailed examination of young brown dwarfs, medium-resolution (R$>$1000) spectra in the near to mid-infrared is required.  These data will lead to distinguishing cloud features from temperature, metallicity or gravity signatures and enable detailed spectral inversion surveys to ground theoretical model predictions (\citep{Burningham17}).  The expected faint near infrared magnitudes ($J$ $<$ 18) of soon to be recovered planetary mass objects means JWST, with its sensitivity and numerous on-board spectrographs covering wavelength ranges where clouds are most influential, will be  critical to young brown dwarf studies in the coming decade. Planned instruments on the Extremely Large Telescopes would also be sensitive enough to open a new window onto cloud physics of super Jupiter mass objects (see for example \citep{Vos19a}).




\begin{figure}[!h]
\vskip -0.1in
 \begin{center}
 \includegraphics[scale=0.3]{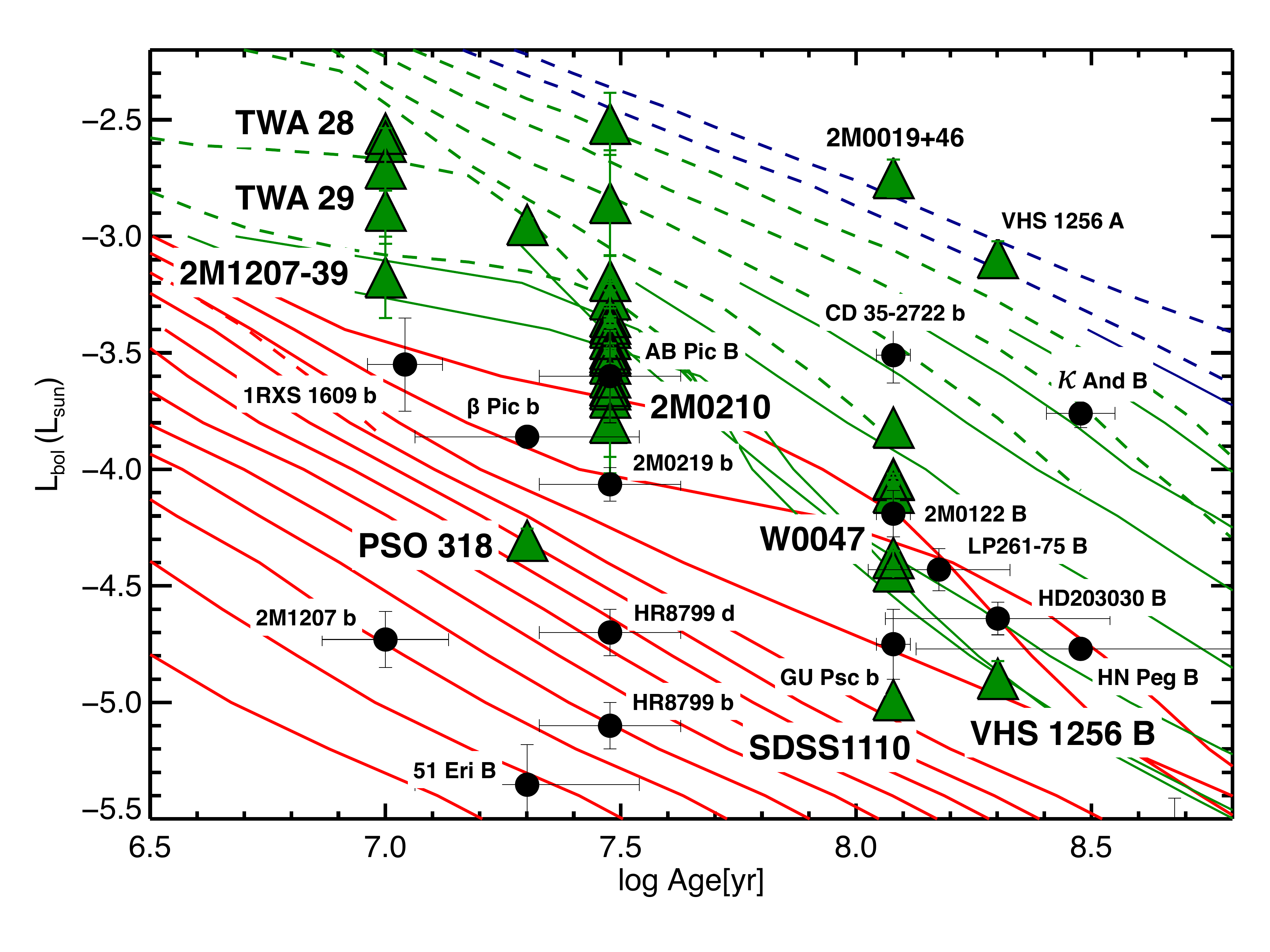}
\end{center}
\vskip -0.5in
\caption{\footnotesize The age versus bolometric luminosity plot with model isochrone tracks at constant mass from \citep{Saumon08} (solid lines) and B\citep{Baraffe15} (dashed lines). We have color coded $<$ 13 M$_{Jup}$ tracks in red, 13 M$_{Jup}$ $<$ M $<$ 75 M$_{Jup}$ tracks in green and $>$ 75 M$_{Jup}$ in blue. Over-plotted are a collection of both the young brown dwarfs and directly imaged exoplanets with measured quantities from \citep{Faherty16}.}
 \label{fig:Lbol}
\end{figure}

\begin{center}
\vskip -0.2in
{\bf 4. Young Brown Dwarfs Put Directly Imaged Exoplanet Data into Context.} 
\end{center} 
\vskip -0.1in
The targeting ground for directly-imaged exoplanets are the exact young associations near the Sun for which we are (and will be) identifying isolated brown dwarfs.  Consequently, isolated planetary-mass brown dwarfs in these groups are critical for placing directly-imaged exoplanets into context. The sources will share initial conditions and subsequent parameters such as age and metallicity. As shown in Figure ~\ref{fig:Lbol} there is already strong overlap in the populations of young brown dwarfs and giant planetary mass companions. Discoveries by instruments such as the Gemini Planet Imager, SPHERE, and Project 1640 have already used the population of young brown dwarfs in moving groups for comparison with newly found planets (e.g. \citep{Macintosh15}, \citep{Bonnefoy16}, \citep{Chilcote17}). 
\medskip\noindent

Isolated young brown dwarfs are equivalent in properties to giant planets (T$_{eff}$, Lbol, Mass, gravity) yet vastly easier to study given that they do not have a bright host star to block in order to detect precious photons.  Consequently studies of the abundances of young brown dwarfs (e.g. C/O ratio) which can be obtained through retrieval studies (see for example \citep{Burningham17}) can be compared to sibling planets to tease out differences in formation mechanisms. Medium resolution spectra in the near and mid infrared can be used to investigate cloud properties and to help unwrap the influence of gravity, metallicity, weather, temperature and mass on the emergent observables. In the coming decade, it is critical that brown dwarf and directly-imaged exoplanet science continue to thrive in context. 


\end{justify}
\pagebreak

\end{document}